\documentclass{svproc}
\usepackage{url}
\usepackage{graphicx}
\usepackage{floatrow}

\begin{document}
\mainmatter              
\title{Investigation of neutron-induced reaction at the Goethe University Frankfurt}
\titlerunning{FRANZ 0.1}  
%
\author{
Ren\'e~Reifarth\inst{1} \and 
Lukas~Bott\inst{1} \and 
Benjamin~Br\"uckner\inst{1} \and 
Ozan~Dogan\inst{1} \and  
Markus~Dworac\inst{1} \and  
Anne~Endres\inst{1} \and  
Philipp~Erbacher\inst{1} \and  
Stefan~Fiebiger\inst{1} \and  
Roman~Gernh\"auser\inst{2} \and  
Kathrin~G\"obel\inst{1} \and  
Fabian~Hebermehl\inst{1} \and  
Tanja~Heftrich\inst{1} \and  
Christoph~Langer\inst{1} \and  
Tanja~Kausch\inst{1} \and  
Nadine~Klapper\inst{1} \and  
Kafa~Khasawneh\inst{1} \and  
Christoph~K\"oppchen\inst{1} \and  
Sabina~Krasilovskaja\inst{1} \and  
Deniz~Kurtulgil\inst{1} \and  
Markus~Reich\inst{1} \and  
Markus~S.~Sch\"offler\inst{1} \and  
Lothar~Ph.~H.~Schmidt\inst{1} \and  
Christian~Schwarz\inst{1} \and  
Zuzana~Slavkovsk\'a\inst{1} \and  
Kurt~E.~Stiebing\inst{1} \and  
Benedikt~Thomas\inst{1} \and  
Meiko~Volknandt\inst{1} \and  
Mario~Weigand\inst{1} \and  
Michael~Wiescher\inst{3}\and 
Patric~Ziel\inst{1}
}
\authorrunning{R. Reifarth et al.} 
%
%
\institute{
Goethe University Frankfurt, Frankfurt, Germany\\
\and
Technical University Munich, Munich, Germany
\and
University of Notre Dame, Notre Dame, IN, USA
\email{reifarth@physik.uni-frankfurt.de}
}

\maketitle              

\begin{abstract}

We present first results and plans for future neutron activation measurements at the Goethe University Frankfurt. The measurements were performed at the Van-de-Graaff accelerator employing the $^{7}$Li(p,n) reaction.
	
\keywords{neutron activation, stellar nucleosynthesis, $\gamma$-detection, charged-particle detection}
\end{abstract}
\section{Introduction}

Neutron-induced reactions are relevant for many astrophysical scenarios \cite{RLK14}. The involved isotopes can be stable as during the s-process or radioactive as during the p-, i- or r-process. The different scenarios are characterized by different temperatures and neutron densities. Direct measurements of the relevant cross section  
are therefore ideally performed for many different energies. The most general method is the time-of-flight method, which typically requires large samples of isotopically enriched material, intense neutron sources and sophisticated detectors.

The activation method alleviates many of these costly requirements at the cost of integral measurements. The Van de Graaff accelerator at the Goethe University Frankfurt provides unpulsed proton beams of up to 20~$\mu$A in the energy regime between 1.5 and 2.5~MeV. This is ideally suited for the production of neutrons via the $^{7}$Li(p,n) reaction. Many different energy spectra can be produced depending on the proton energy, the thickness of the lithium layer and the position of the irradiated sample.

\section{Activation technique}

\begin{figure}
\floatbox[{\capbeside\thisfloatsetup{capbesideposition={right,top},capbesidewidth=4cm}}]{figure}[\FBwidth]
{\caption{A typical activation setup consists of a continuous neutron source, a sample positioned very close to the neutron source and a separate setup to detect to decay of freshly produced, radioactive nuclei.}\label{activation_setup_new}}
{\includegraphics[width=0.55\textwidth]{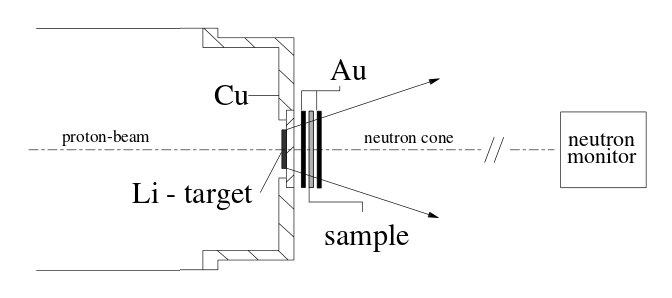}}
\end{figure}

To obtain stellar cross sections from an activation experiment, the neutron spectrum should 
ideally correspond to the thermal spectrum at the respective $s$-process site \cite{REF18}. The $^7$Li(p,n)$^7$Be reaction, which represents the most prolific neutron source 
at low energy accelerators fulfills this requirement almost perfectly \cite{RaK88,LKM12,FFK12}. 
Adjusting the proton energy at $E_p=1912$~keV, 30~keV above the reaction threshold, yields 
a neutron spectrum with an energy dependence close to a Maxwellian distribution corresponding to 
$k_BT=25$~keV almost perfectly mimicking the situation during He shell flashes in AGB stars. 
A typical activation setup is depicted in Fig.~\ref{activation_setup_new}.

The neutron spectrum can be significantly altered, if the proton energy, the proton-energy distribution, the thickness of the 
lithium target, or the angular coverage of the neutron field by the sample is modified \cite{RHF08,RHK09}. The neutron flux is typically determined using a reference sample of the same shape up- and downstream of the sample. 

\section{First measurements}
The upcoming neutron facility FRANZ in Frankfurt (Germany) \cite{RCH09} will be based on the $^7$Li(p,n) reaction to produce neutrons by upgrading the proton source as well as high 
current lithium targets. In a first step, the  Van-de-Graaff accelerator at the department of physics at the Goethe University Frankfurt was used to perform measurements of neutron-induced reactions.

\subsection{Detection of $\gamma$-activities}

\begin{center}
\begin{figure}
\centering
\includegraphics[width=0.50\textwidth]{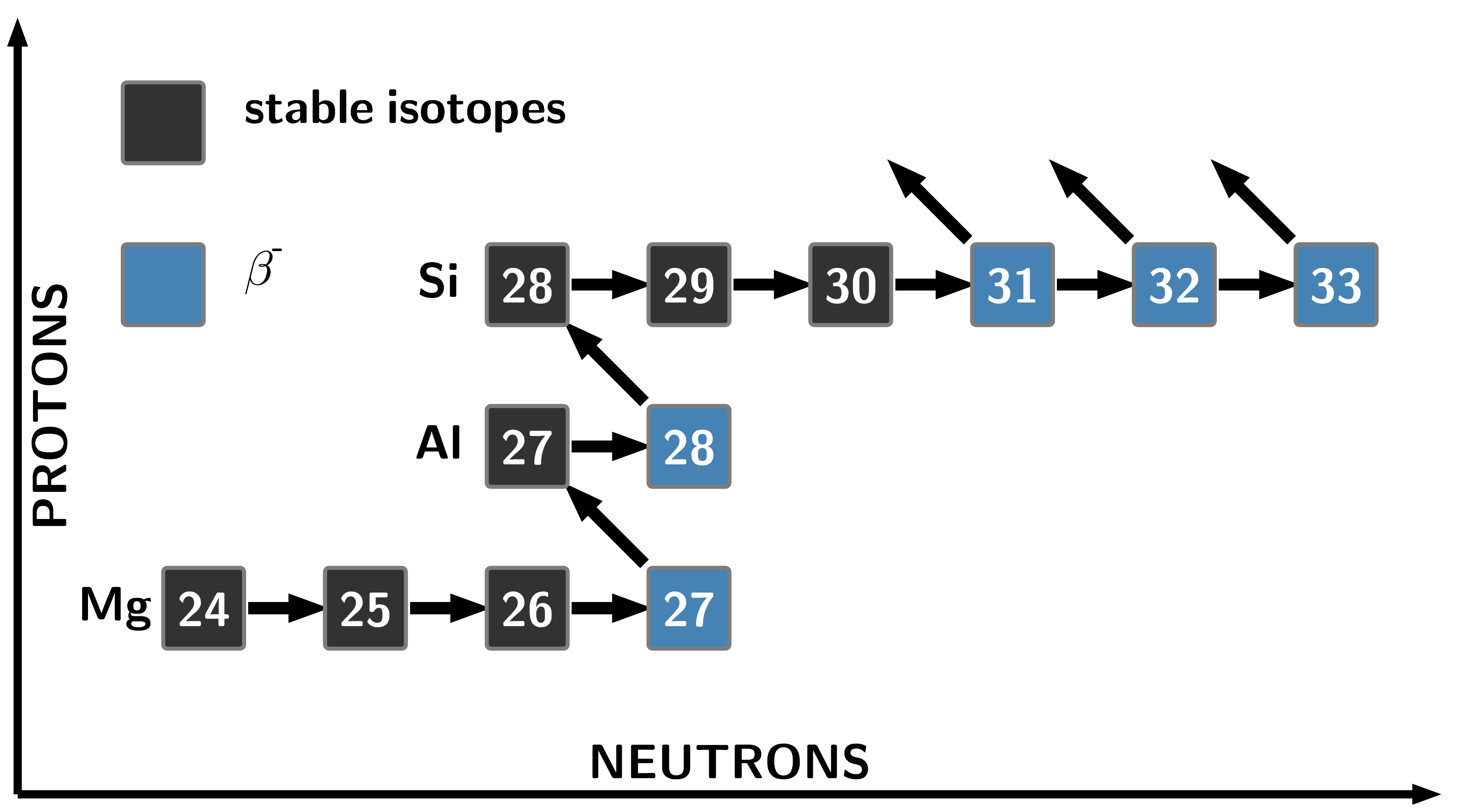}
\includegraphics[width=0.45\textwidth]{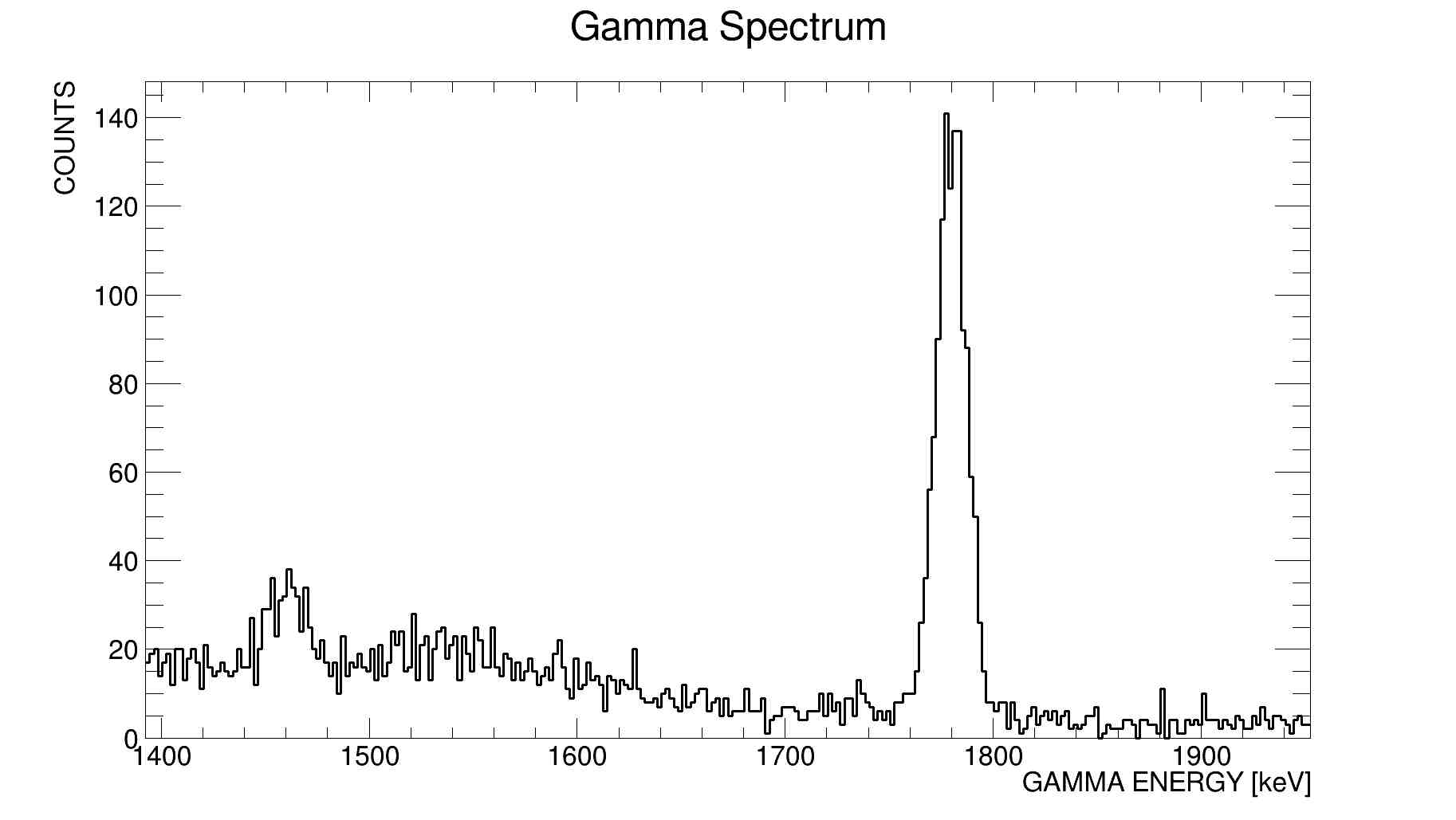}
   \caption{The s-process reaction network around $^{27}$Al (left) and $\gamma$-spectrum resulting from the 2.2~min decay of $^{28}$Al following the activation of natural aluminum (right).
   \label{al_and_more} }
\end{figure}
\end{center}

\begin{center}
\begin{figure}
\centering
\includegraphics[width=0.5\textwidth]{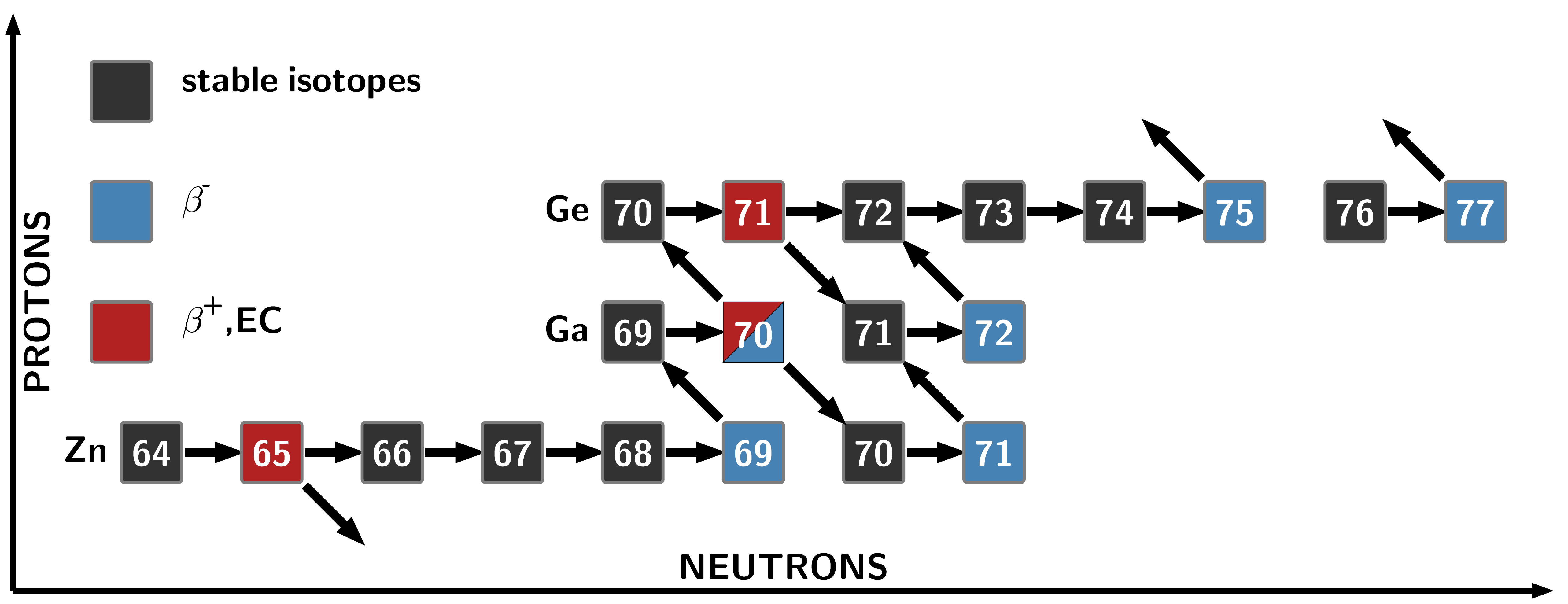}
\includegraphics[width=0.45\textwidth]{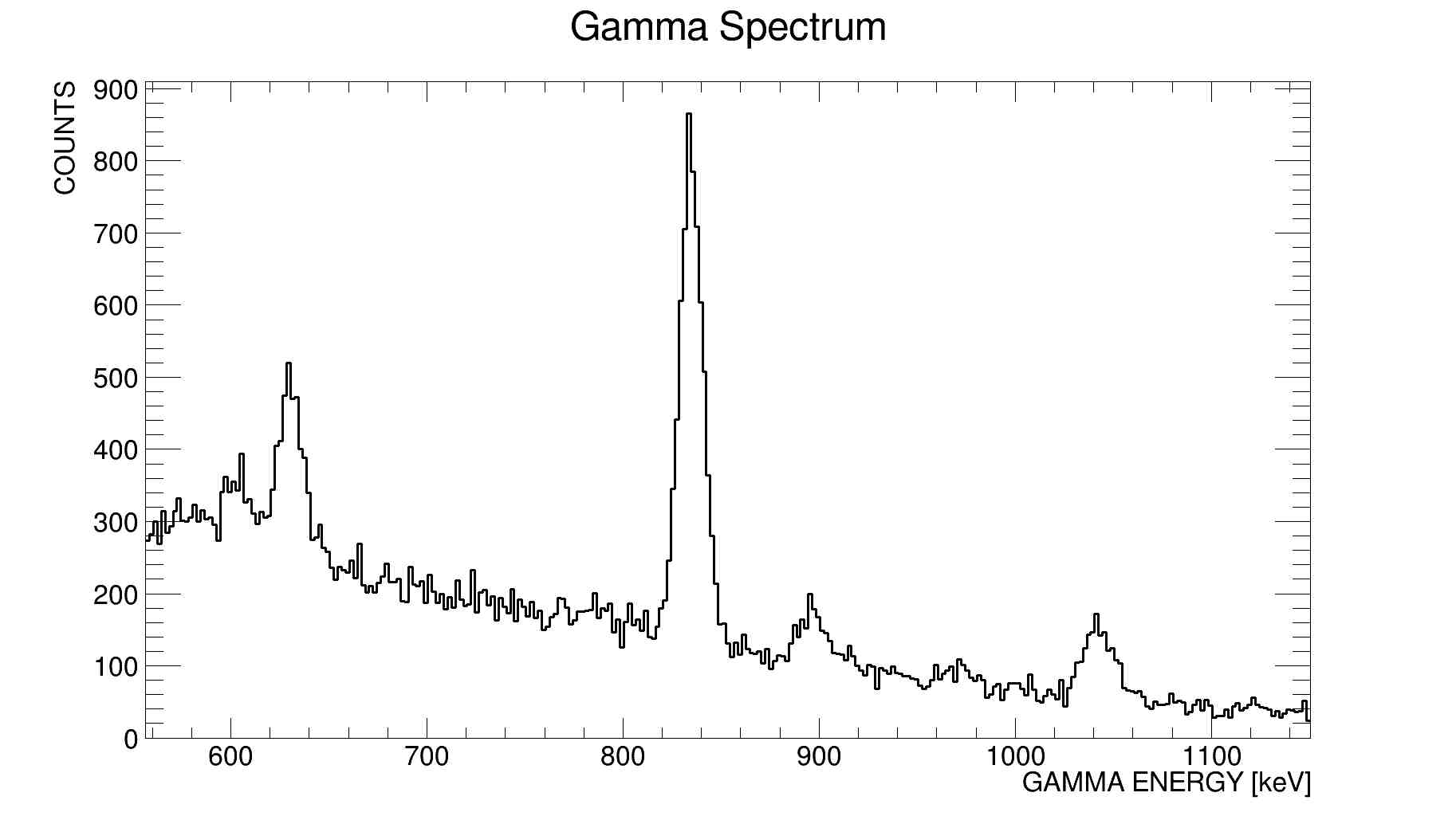}
   \caption{The s-process reaction network around gallium (left) and $\gamma$-spectrum resulting from the decay of $^{70,72}$Ga following the activation of natural gallium (right). 
   \label{ga_and_more} }
\end{figure}
\end{center}

\begin{center}
\begin{figure}
\centering
\includegraphics[width=0.50\textwidth]{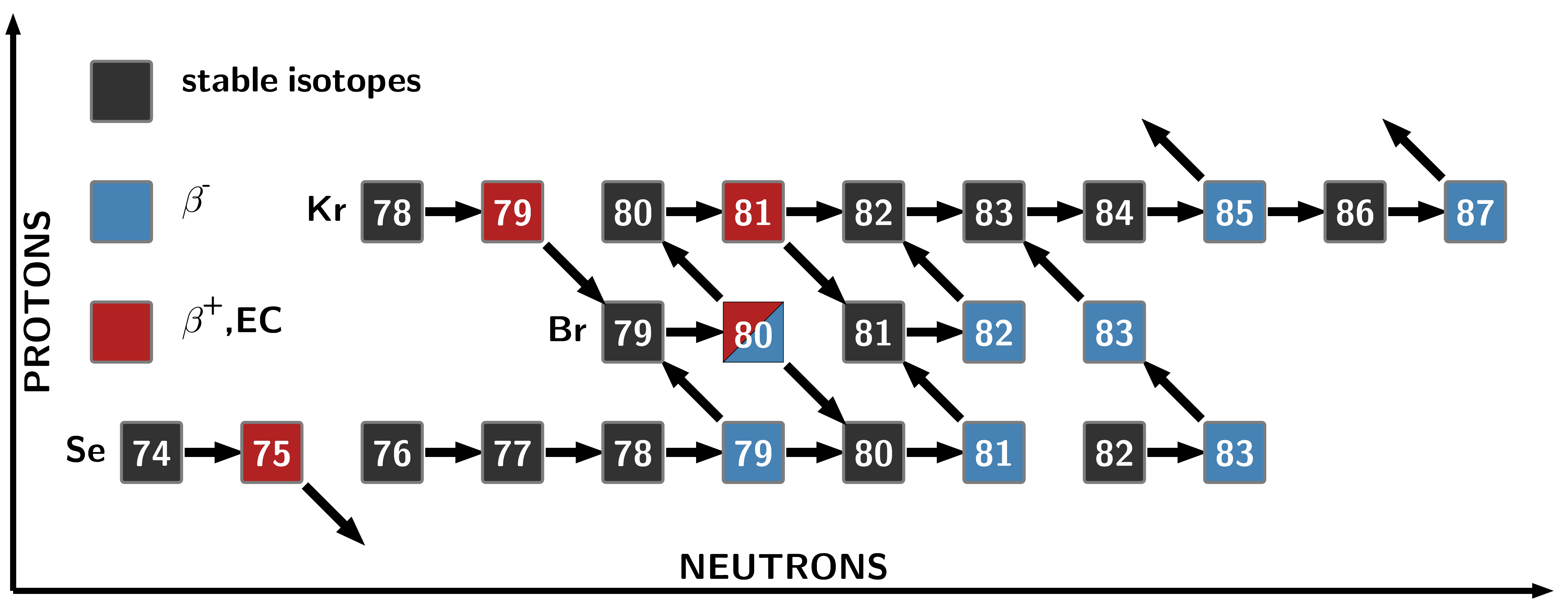}
\includegraphics[width=0.45\textwidth]{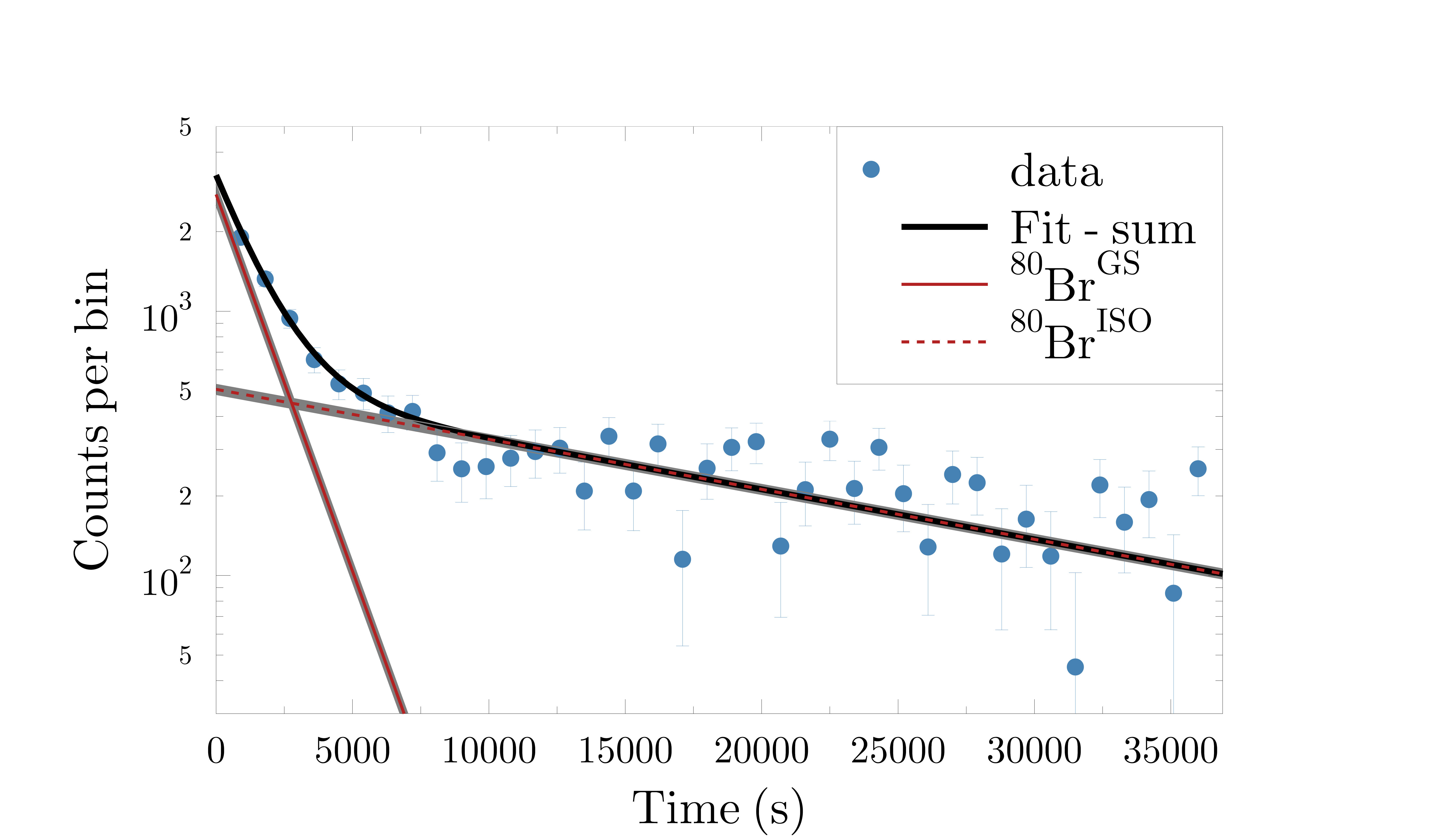}
   \caption{The s-process reaction network around bromine (left) and the time dependence of the 511~keV $\gamma$-line resulting from the decay of $^{80}$Br$^\mathrm{GS}$ following the activation of natural bromine (right).
   \label{br_and_more} }
\end{figure}
\end{center}

We performed activations of natural samples of aluminum, potassium chloride, gallium and potassium bromide, see Figs~\ref{al_and_more}-\ref{br_and_more}. While recent activation measurements for the isotopes $^{27}$Al~\cite{HKU06}, $^{41}$K~\cite{HPU16}, and $^{79,81}$Br \cite{HKU08} exist, no activation-based data are available for $^{37}$Cl, $^{69,71}$Ga as well as for the population of the isomeric state of $^{82}$Br. The preliminary results shown in Figs~\ref{al_and_more}-\ref{br_and_more} indicate that the activations were successful and we anticipate final results with 5-10\% uncertainty. 

\subsection{Detection of $\alpha$-activities}

\begin{figure}
\floatbox[{\capbeside\thisfloatsetup{capbesideposition={right,top},capbesidewidth=4cm}}]{figure}[\FBwidth]
{\caption{Sketch of the NICE-setup used to measure the $^7$Li(n,$\gamma$) cross section. The NICE-detector (Neutron-Induced Charged particle Emission) is based on a thin scintillation foil read out by photo-multiplier tubes mounted on flat side of the foil.}\label{nice_setup}}
{\includegraphics[width=0.5\textwidth]{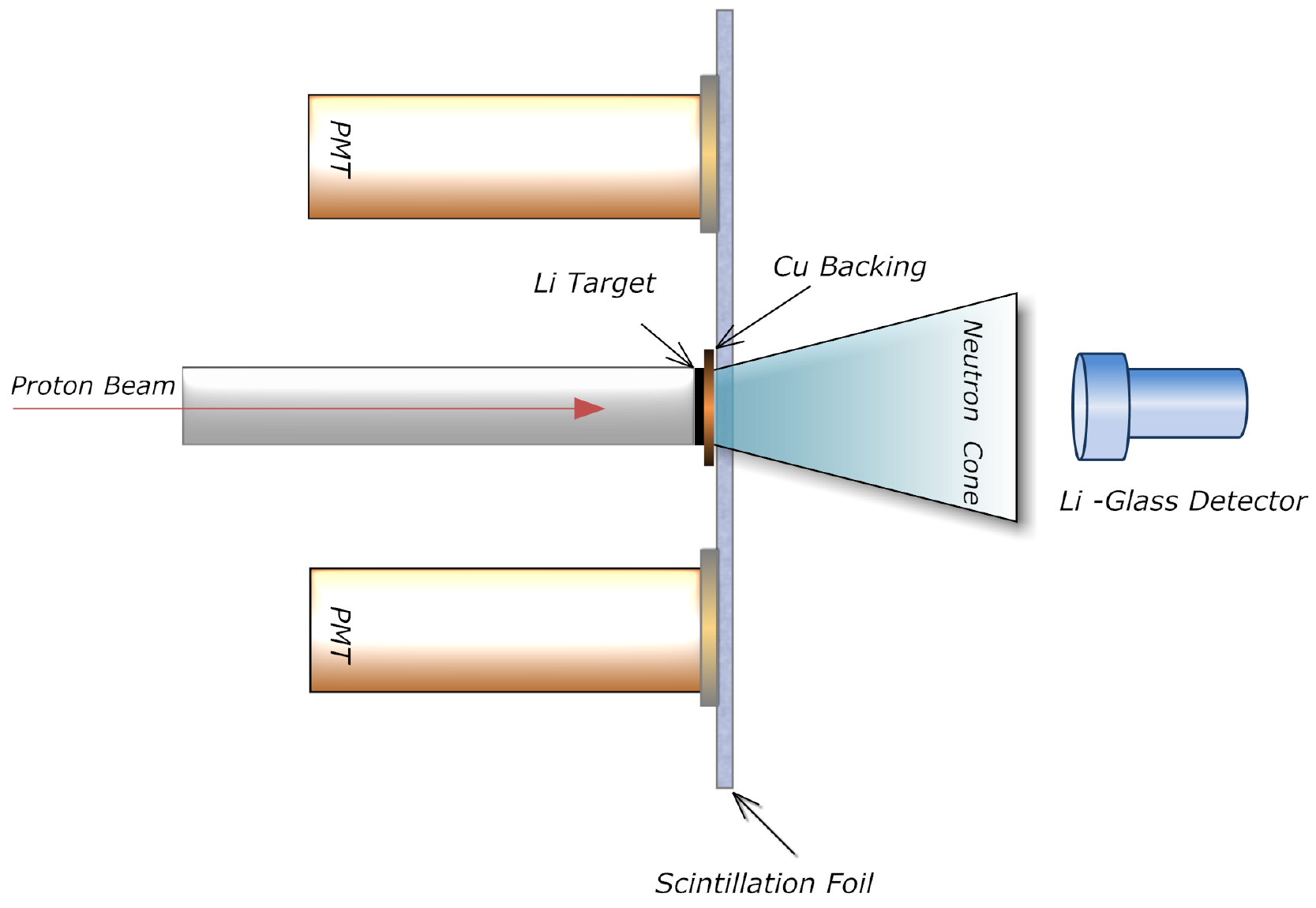}}
\end{figure}

A new detection system for charged particles was successfully tested for the first time, see Fig.~\ref{nice_setup}. This setup allows the investigation of neutron-induced reactions with charged particles in the exit channel. This includes (n,$\alpha$), (n,p), and (n,fission) reactions. A first test was performed on the $^7$Li(n,$\gamma$)$^8$Li($\beta^-$)$^8$Be($\alpha$)$\alpha$ reaction. This reaction was investigated in the past with ionization chambers, \cite{HKW98}. For this purpose, the proton beam from the accelerator was periodically directed onto the neutron production target for 2~s and then deflected onto a beam stop for 2~s to observe the 0.8~s decay of $^8$Li followed by the prompt $\alpha$-decay of $^8$Be.

\section{Outlook}

In the near-term future we plan to perform activation measurements with different neutron spectra to constrain the stellar reaction rates at different temperatures. As soon as the RFQ-based accelerator is operational, we will perform activation as well as time-of-flight measurements on radioactive isotopes.

\section*{Acknowledgments}

This research has received funding from the European Research Council under the European Unions's Seventh Framework Programme (FP/2007-2013) / ERC Grant Agreement n. 615126 and from the Deutsche Forschungsgemeinschaft RE-3461/4-1 and RE 3461/4-1.

%
%
\bibliographystyle{spphys}
\bibliography{/home/reifarth/Texte/paper/refbib}
\end{document}